\documentclass[11pt,a4paper]{article}
\usepackage{jheppub}
\usepackage{amsmath,amsthm,amssymb}
\usepackage[dvipsnames,svgnames,x11names]{xcolor}
\usepackage[normalem]{ulem}
\usepackage{braket}
\usepackage{subcaption}
\usepackage{mathtools}
\usepackage{hyperref}
\hypersetup{
    colorlinks=true,                 
    linkcolor=DeepSkyBlue4,                 
    linktoc=all,                     
    citecolor=SeaGreen,                 
    filecolor=pink,                  
    urlcolor=OliveDrab                 
}

\newcommand{\del}{\partial}

\DeclareMathOperator{\sech}{sech}

\newcommand{\Rmnum}[1]{\expandafter\@slowromancap\romannumeral #1@}

\newcommand{\nn}{\nonumber}

\newcommand{\be}{\begin{equation}}
\newcommand{\ee}{\end{equation}}

\newcommand{\lB}{\left [}
\newcommand{\rB}{\right ]}
\newcommand{\lb}{\left (}
\newcommand{\rb}{\right )}

\renewcommand{\a}{\alpha}	
\newcommand{\e}{\epsilon}

\newcommand{\cF}{\mathcal{F}}

\newcommand{\cQ}{\mathcal{Q}}

\title{Quantum null energy condition in quenched 2d CFTs}
\begin{abstract}{The quantum null energy condition (QNEC) is a lower bound on the expectation value of the null-null component of the energy-momentum tensor in terms of null variations of the entanglement entropy. A stronger version of the QNEC (the primary QNEC) is expected to hold in $1+1$ dimensional conformal field theories (CFT). QNEC has been shown to impose non-trivial quantum thermodynamic restrictions on irreversible entropy production in quenches in $1+1$ dimensional holographic CFTs. It is therefore natural to study if QNEC imposes similar bounds in other quench setups. In this paper we study QNEC in the Calabrese-Cardy global and local joining quenches using standard CFT techniques. In the global quench we show that the primary QNEC must hold at sufficiently early times and find that it imposes bounds on the four point correlators of twist fields in a boundary state. This is a constraint on the set of boundary states that {satisfy} the primary QNEC. Furthermore, we find that a violation of the primary QNEC implies a violation of the averaged null energy condition (ANEC) in a conformally transformed frame. In the local quench we find similar bounds on four point correlators from { both the primary and the usual} QNEC.}
\end{abstract}

\author{Tanay Kibe}
\emailAdd{tanay.kibe@wits.ac.za}
\author{and Pratik Roy}%
\emailAdd{roy.pratik92@gmail.com }
\affiliation{National Institute for Theoretical and Computational Sciences,
School of Physics\\ and Mandelstam Institute for Theoretical Physics,
University of the Witwatersrand,
Wits, 2050, South Africa}%

\begin{document}
\maketitle


\section{Introduction}
Quantum information theory has played a crucial role in unravelling many aspects of the structure of quantum field theory and quantum gravity \cite{Faulkner:2022mlp}. One interesting recent development has been the quantum null energy condition (QNEC), a quasi-local lower bound on the energy-momentum tensor in Poincar\'{e} invariant quantum field theories (QFT).  A very general information-theoretic argument for QNEC was presented in \cite{Wall:2017blw}. The argument is essentially captured by Landauer's proclamation that information is physical \cite{Landauer:1961,Landauer:1991}, i.e., some minimum energy is required to store (erase) a given amount of information \cite{Reeb_2014}. 

QNEC was first conjectured as a special non-gravitational ($G_N$-independent) limit of the Quantum Focusing Conjecture \cite{Bousso:2015mna}. It has been proven for free quantum field theories (QFTs) \cite{Bousso:2015wca,Malik:2019dpg}, holographic QFTs with AdS duals \cite{Koeller:2015qmn},
and for general Poincar\'e-invariant QFTs \cite{Balakrishnan:2017bjg}. More rigorous proofs  using half-sided modular inclusion properties of operator algebras have also been provided \cite{Ceyhan:2018zfg, Hollands:2025glm}.

It is interesting to ask if QNEC imposes non-trivial constraints on the dynamics of out-of-equilibrium many-body systems described by QFTs. In earlier work \cite{Kibe:2021qjy,Banerjee:2022dgv,Kibe:2024icu}, we found that non-violation of the QNEC imposes upper and lower bounds on irreversible entropy production after a global quench in holographic conformal field theories (CFT). These bounds were argued to be a non-trivial generalization of the classical second law of thermodynamics and {were already} expected from general quantum information theory arguments \cite{PhysRevLett.105.170402}. It is therefore natural to investigate QNEC in a variety of other quenches and examine the bounds imposed on these quenches.

This paper describes results of calculations of the QNEC for quenches in $1+1$ dimensional CFTs. The QNEC is \cite{Bousso:2015mna}
\begin{equation}
    Q_\pm \coloneqq 2 \pi \langle T_{\pm\pm}\rangle - \partial_\pm^2 S  \geq 0,
     \label{Eq:qnecweak}
\end{equation}
where $\langle T_{\pm\pm}\rangle$ is the expectation value of the two non-zero components of the energy-momentum tensor of the CFT at an arbitrary spacetime point $p$ with $+(-)$ denoting the future directed right (left) moving null directions, $S$ is the entanglement entropy of any spacelike interval with one end-point at $p$, and $\partial_\pm S$ is the rate of change of entropy when the end-point $p$ is deformed along the $\pm$ directions while the other end-point is kept fixed. 

In $1+1$-dimensional CFTs, a stronger QNEC,
\begin{equation}        \label{eq:QNEC-def}
   \mathcal{Q}_\pm \coloneq 2 \pi \langle T_{\pm\pm}\rangle - \left(\partial_\pm^2 S + \frac{6}{c}\left(\partial_\pm S\right)^2\right) \geq 0,
\end{equation}
is implied by the inequality \eqref{Eq:qnecweak} \cite{Bousso:2015mna, Wall:2011kb}. The argument for this stronger inequality is as follows. The quantity $\mathcal{Q}_\pm$ transforms like a conformal primary and one can transform to a conformal frame where the $\partial_+S$ term vanishes. The transformation {$y\to f(y)$ that achieves this can be found by solving}
\begin{equation}
    S'(f) = f'(y)^{-2}\left(\frac{c}{12}S'(y)+ \frac{f''(y)}{f'(y)}\right)=0.
    \label{Eq:sprimezero1}
\end{equation}
The QNEC \eqref{Eq:qnecweak} must hold in this new frame if the assumptions in the proof by Ceyhan and Faulkner (CF) \cite{Ceyhan:2018zfg} hold. Since $\mathcal{Q}_\pm$ transforms like a primary, we will henceforth refer to \eqref{eq:QNEC-def} as primary QNEC.\footnote{This name was suggested by Thomas Faulkner.} We will usually refer to $Q_\pm$ in \eqref{Eq:qnecweak} as the \textit{weak}  QNEC in the rest of paper.

In this paper, we investigate both the primary and weak QNECs in a global \cite{Calabrese:2005in} and a local joining \cite{Calabrese:2007mtj} quench using standard CFT techniques. In both the quenches, we find that the primary QNEC imposes non-trivial bounds on four-point correlators of the theory. The details of the results depend on the quench protocol. For the global quench, we show that one of the primary QNECs must hold, and that it imposes non-trivial bounds on the dynamics. The weak QNEC is satisfied at all times and for all entangling regions in the global quench. For the local quench, we are unable to satisfactorily demonstrate that the primary QNEC holds. However, in this case we find non-trivial bounds from the weak QNEC which approach the bounds from the primary QNEC in a specific limit.

An outline of the paper follows. In Section \ref{sec:global}, we study the global quench setup. We provide a somewhat detailed overview of the computation of entanglement entropy for this case since the calculations are very similar in the other case. 
We then obtain the energy-momentum tensor using the anomalous transformation of the vanishing energy-momentum tensor on the plane. 
With the entanglement entropy and energy-momentum tensor known, QNEC is derived by just taking derivatives. We then proceed to check if the assumptions of the CF proof are satisfied, in particular we check for the finiteness of the averaged null energy (ANE) operator that moves only one of the end-points of the entangling region. In Section \ref{sec:local}, we follow exactly the same path for the local joining quench.  We conclude with discussions about our results and their implications in Section \ref{sec:discussion}.


\section{Global quench}     \label{sec:global}
\begin{figure}
    \centering
    \includegraphics[scale=0.65]{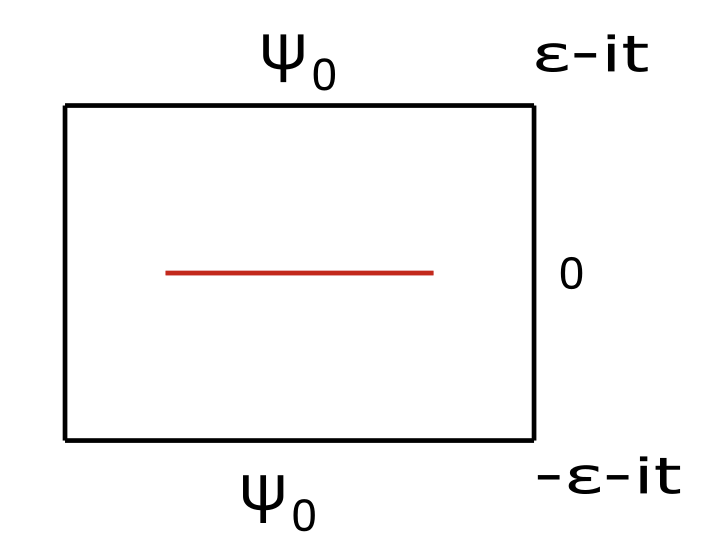}
    \caption{The strip geometry for the Euclidean path integral that prepares the density matrix at a time $t$ after the global quench. The vertical axis is imaginary time. $\ket{\psi_0}$ is the boundary condition for the path integral. The red line indicates a slit in the strip along the interval whose density matrix is being constructed. This figure is adapted from  \cite{Calabrese:2005in}.}
    \label{fig:global-setup}
\end{figure}

We first consider the global quenches in $1+1$ dimensional CFTs described in \cite{Calabrese:2005in}, where the Hamiltonian of the system is changed suddenly, for example by a sudden change in a coupling constant. The system is prepared in a translation invariant eigenstate $
\ket{\psi_0}$ of the Hamiltonian $H_0$. The system is quenched at time $t=0$ by changing the Hamiltonian to $H$, which is assumed to be critical and described by a conformal field theory. The short distance divergences are regulated by modifying the state as $e^{-\epsilon H} \ket{\psi_0}$. Under renormalization group flow the translation invariant state $\ket{\psi_0}$ flows to a conformally invariant boundary state $\ket{B}$. Therefore, for studying the post-quench dynamics for scales much larger than $\epsilon$ one can replace $\ket{\psi_0}$ with $\ket{B}$.  The density matrix for an equal time interval can be prepared using an analytically continued Euclidean path integral on a strip geometry, as shown in Fig. \ref{fig:global-setup}

\subsection{Entanglement entropy}
The R\`{e}nyi entropies for the post-quench state can be calculated using the prescription from \cite{Calabrese:2005in,Calabrese:2004eu}. The $n^{th}$ R\`{e}nyi entropy of a single interval $A$ is given by a twist anti-twist field correlator on the upper half plane (UHP) as
\begin{align}\label{Eq:renyicorr}
        S^{(n)}_A={\rm Tr}(\rho^n_A) &= a(n)\braket{\Phi_n(z_1) \Phi_{-n}(z_2)}_{\text{UHP}} \\ &
        = {\left( |z_1-z_2||\bar{z}_1-\bar{z}_2| |\eta|\right)^{-2n \Delta_n}} \tilde{\mathcal{F}}_n(\eta) \nonumber,
    \end{align}
where $z_{1,2}$ are the end-points of the interval, and $a(n)$ is a coordinate independent theory-dependent proportionality constant. We have absorbed $a(n)$ into the definition of $\tilde{\mathcal{F}}_n$ in the second equality above. $\Phi_{n,-n}$ are twist fields with scaling dimension $\Delta_n = \frac{c}{24}(1-\frac{1}{n^2})$ and $\tilde{\mathcal{F}}_n$ is a function of the conformal cross ratio 
\[
\eta = \frac{(z_1-\bar{z}_1)(z_2-\bar{z}_2)}{(z_1-\bar{z}_2)(z_2-\bar{z}_1)}\, ,
\]
and depends on the full operator content of the theory.\footnote{The tilde is used to be consistent with the notation in \cite{Calabrese:2009qy} and to distinguish $\tilde{\mathcal{F}}_n$ from the function $\mathcal{F}_n$ which appears in the four point correlator of primaries on a plane.} For $n=1$ we have
\begin{equation}
    {\rm Tr}(\rho_A) = \tilde{\mathcal{F}}_1(\eta),
\end{equation}
and the normalization of the density matrix fixes
\begin{equation}
    \tilde{\mathcal{F}}_1(\eta)=1.
\end{equation}
We note that $\tilde{\mathcal{F}}_n(\eta)\approx 1$ when $\eta \approx 0, 1$ \cite{Calabrese:2009qy}. 
The entanglement entropy is
 \begin{equation}
        S_A =-\partial_nS^{(n)}_A \Big\vert_{n=1}.
\end{equation}
This can be easily calculated in the UHP coordinates to be
\begin{equation}            \label{eq:EE-global-UHP}
    S_A = -\frac c6\log \lB \frac1{(z_1 - z_2) (\bar z_1 - \bar z_2)} + \frac1{(z_1 - \bar z_1) (\bar z_2 - z_2)} \rB - \partial_n\tilde{\mathcal{F}}_n(\eta)\vert_{n=1} \ .
\end{equation}
     
The reduced density matrix after the global quench is defined via the Eucliden path integral on a strip geometry with width $2\epsilon$ and a branch cut on the interval $A$ of interest.
The transformation
\begin{equation}
     w = \frac{2 \epsilon}{\pi} \log z,
\end{equation}
maps the UHP to the strip. Using the transformation of the correlator of primary fields \eqref{Eq:renyicorr} to map it to the strip and analytically continuing the Euclidean time to $\tau = \epsilon +i t$, we can readily express the R\`{e}nyi entropies in terms of the Minkowski coordinates of the end-points of the interval. After analytic continuation we have $w=-x^-$ and $\bar{w}=x^+$. 
The entanglement entropy (\ref{eq:EE-global-UHP}) can be expressed in terms of the light-cone coordinates $x^\pm_{1,2}$ of the end-points of the interval as
\begin{align}
	S_A &= -\frac c6\log\lB - \frac{\pi^2\cosh\lb\frac{\pi(x_2^+ + x_1^-)}{4\e}\rb \cosh\lb\frac{\pi(x_1^+ + x_2^-)}{4\e}\rb \sech\lb\frac{\pi(x_1^+ + x_1^-)}{4\e}\rb\sech\lb\frac{\pi(x_2^+ + x_2^-)}{4\e}\rb}{16 \epsilon^2\sinh\lb\frac{\pi(x_2^+ - x_1^+)}{4\e}\rb \sinh\lb\frac{\pi(x_2^- - x_1^-)}{4\e}\rb} \rB \\& \qquad - \partial_n\tilde{\mathcal{F}}_n(\eta)\vert_{n=1}\nonumber,
\end{align}
where the cross ratio is
\begin{equation}
    \eta= \frac{\left(e^{\frac{\pi  (x_1^{-}+x_1^{+})}{2 \epsilon }}+1\right) \left(e^{\frac{\pi  (x_2^{-}+x_2^{+})}{2 \epsilon }}+1\right)}{e^{\frac{\pi  (x_1^{-}+x_2^{-}+x_1^{+}+x_2^{+})}{2 \epsilon }}+e^{\frac{\pi  (x_1^{-}+x_2^{+})}{2 \epsilon }}+e^{\frac{\pi  (x_2^{-}+x_1^{+})}{2 \epsilon }}+1}\, .
\end{equation}
We note however that this form of the entropy should not be used to calculate the derivatives that appear in $Q_\pm,\cQ_\pm$. We need to differentiate first in the UHP coordinates and only analytically continue when we have obtained the quantity of interest.\footnote{We have checked that this subtlety is not important for the global quench. However, this will be important for the local quench.}

Due to translation invariance, there is no loss of generality in choosing the end-points to be
\begin{equation}			\label{eq_end_pts_global_quench}
	x_1^+ = t - \frac l2, \quad x_1^-  = t+ \frac l2, \qquad x_2^+ = t + \frac l2, \quad x_2^- = t - \frac l2.
\end{equation}
For $\epsilon \ll l,t$  
 \begin{align}
     &\eta\approx e^{-\frac{\pi(l-2t)}{2\epsilon}}\sim0  \qquad\quad\ \ \text{when } t<\frac l2\, , \nonumber\\
      &\eta\approx 1-e^{-\frac{\pi(2t-l)}{2\epsilon}}\sim 1  \qquad\text{when } t>\frac l2\, .
 \end{align}
These imply
\begin{align}
    \tilde{\mathcal{F}}_{n}(\eta)&\approx 1+ \eta \tilde{\mathcal{F}}_n'(0),\qquad\qquad\ \,\text{for } l>2t,\\
    \tilde{\mathcal{F}}_{n}(\eta)&\approx 1+ (\eta-1) \tilde{\mathcal{F}}_n'(1) ,\qquad\text{for } l<2t,
\end{align}
where prime denotes a derivative with respect to $\eta$. We have not kept higher order terms in the above expansion of $\tilde{\mathcal F}$ since we find that they contribute at sub-leading order in the QNEC.
\subsection*{Energy-momentum tensor}
The energy-momentum tensor can be calculated using the Schwarzian derivative of the map from the UHP vacuum to the strip and using the fact that the expectation value of the energy-momentum tensor vanishes in the UHP vacuum. The transformation of the energy-momentum tensor is
\begin{equation}
    \braket{T(z)}=\lb\frac{\partial z}{\partial w}\rb^{-2}\lb\braket{T(w)}+\frac{c}{24\pi}{\rm Sch}(z,w)\rb,
\end{equation}
where $${\rm Sch}(z,w)=\frac{z'''(w)}{z''(w)}-\frac{3}{2}\lb\frac{z''(w)}{z'(w)}\rb^2$$
is the Schwarzian derivative. We obtain $\braket{T(w)}$ by setting $\braket{T(z)}=0$ in the above expression.
The energy-momentum tensor in the lightcone coordinates is
\begin{equation}
       \braket{T_{++}(x^+)}=\braket{T_{--}(x^-)}=\frac{c \pi}{192 \epsilon^2} \, .
    \end{equation}

\subsection{QNEC}
Using the energy-momentum tensor and the expression for the entanglement entropy one can calculate the left hand side of the QNEC inequalities $\mathcal{Q}_\pm$ (\ref{eq:QNEC-def}) at both end-points in an expansion with $\epsilon \ll t,l$. When $t<\frac l2$, we have
{
\begin{align}
    \mathcal{Q_+}(1) = \mathcal{Q}_-(2) &\approx
    \frac{\pi^2}{2 \epsilon^2}e^{-\frac{\pi(l-2t)}{2 \epsilon}} \lb \frac{c}{6}    +  \partial_n\tilde{\mathcal{F}}'_n(0)\vert_{n=1} \rb
    ,\\ 
    \mathcal{Q_+}(2)= \mathcal{Q_-}(1) &\approx \frac{\pi^2}{2 \epsilon^2}e^{-\frac{\pi(l+2t)}{2 \epsilon}} \lb \frac{c}{6} + \partial_n\tilde{\mathcal{F}}'_n(0)\vert_{n=1} \rb ,
    \label{Eq:qnecglobal1}
\end{align}}
and for $t>\frac l2$, we have
{
\begin{align}
    \mathcal{Q_+}(1) = \mathcal{Q}_-(2) &\approx \frac{\pi^2}{2 \epsilon^2}e^{\frac{\pi(l-2t)}{2 \epsilon}} \lb \frac{c}{6} - \partial_n\tilde{\mathcal{F}}'_n(1)\vert_{n=1} \rb, \\
    \mathcal{Q_+}(2)= \mathcal{Q_-}(1) &\approx \frac{\pi^2}{2 \epsilon^2}e^{-\frac{\pi(l+2t)}{2 \epsilon}} \lb \frac{c}{6} - \partial_n\tilde{\mathcal{F}}'_n(1)\vert_{n=1} \rb.
\end{align}
}
Here, the $1(2)$ in parentheses after $\mathcal{Q}_\pm$ indicate that the null variations of the entanglement entropy are computed at the first (second) end-point of the interval. Demanding that $\mathcal{Q}_\pm(1,2)\geq0$ implies that
\begin{equation}
    \frac{c}{6}+\partial_n\tilde{\mathcal{F}}'_n(0)\vert_{n=1}\geq 0, \quad \text{and } \quad  \frac{c}{6}-\partial_n\tilde{\mathcal{F}}'_n(1)\vert_{n=1}\geq 0.
    \label{Eq:qnecboundglobal}
\end{equation}
We have only kept the leading terms in the above QNEC expansions. The sub-leading terms are exponentially suppressed relative to the leading term. We also find that the weak QNEC \eqref{Eq:qnecweak} is always satisfied in this case and is the following at leading order:
\begin{equation}
    Q_\pm(1,2) \approx \langle T_{\pm\pm} \rangle = \frac{c \pi}{192 \epsilon^2} \,.
\end{equation}
The sub-leading terms in $Q_\pm$ are exponentially suppressed relative to the leading term.

Note that a naive calculation that ignores the contribution of the theory-dependent function $\tilde{\mathcal{F}}$ would lead to QNEC inequalities that are identically satisfied. The naive QNEC answer can be obtained by setting $\tilde{\mathcal{F}}_n$ to zero in the above expressions. Although the contribution of $\tilde{\mathcal{F}}_n$ to the entropy is sub-leading, the null derivatives of this function contribute at leading order in the QNEC.\footnote{Heuristically, for $t<l/2$ since $\eta \approx e^{-\frac{\pi(l-2t)}{2\epsilon}}$ taking null derivatives of $\eta$ pulls out factors of $\epsilon^{-1}$, which lead to a contribution to the QNEC at leading order.} Finally, we note that the QNEC is saturated for a semi-infinite interval, i.e. when $l \to \infty$ in \eqref{Eq:qnecglobal1}, which is the answer obtained in \cite{Mezei:2019sla}. 

\subsection{Averaged null energy}\label{ssec:checkaneglobal}

The general proof of QNEC \cite{Ceyhan:2018zfg} assumes that the state in which we compute the QNEC has finite relative entropy with respect to the vacuum, and that the averaged null energy (ANE) of the state,
\begin{equation}
    \braket{\mathcal{P}} = \int_{-\infty}^{\infty} \braket{T_{\pm\pm}}dx^\pm\, ,
    \label{Eq:ANE}
\end{equation}
is finite. Note that the ANE operator generates translations of the end-point of the half line along the $\pm$ direction. The results we have obtained for the global quench do not depend on the details of the Hamiltonians $H_{0,1}$ and the eigenstate $\ket{\psi_0}$. We expect that one
 can engineer the quench such that the $t=0$ boundary state has finite relative entropy with respect to the CFT vacuum, for example if the two Hamiltonians $H_{0,1}$ are infinitesimally close. 

In order to check the finiteness of the ANE, we must transform to the conformal frame where the first variation of the entropy vanishes and the CF proof can be applied. The conformal transformation that achieves this is \eqref{Eq:sprimezero1}
\begin{equation}
    y\to f(y), \quad f'(y) =\alpha \exp\left(-\frac{12}{c}S(y)\right),
    \label{Eq:sprimezero}
\end{equation}
where $\alpha$ is an arbitrary constant, which we choose to be positive, and $S(y)$ is the entanglement entropy in the $y$ frame.

The entanglement entropy as a function of $x^-$ at the first end-point is 
\begin{align}
	S(x^-) &= -\frac c6\log\lB - \frac{\pi^2\cosh\lb\frac{\pi(x_2^+ + x^-)}{4\e}\rb \cosh\lb\frac{\pi(x_1^+ + x_2^-)}{4\e}\rb \sech\lb\frac{\pi(x_1^+ + x^-)}{4\e}\rb\sech\lb\frac{\pi(x_2^+ + x_2^-)}{4\e}\rb}{16 \epsilon^2\sinh\lb\frac{\pi(x_2^+ - x_1^+)}{4\e}\rb \sinh\lb\frac{\pi(x_2^- - x^-)}{4\e}\rb} \rB \\& \qquad - \partial_n\tilde{\mathcal{F}}_n(\eta)\vert_{n=1}\nonumber,
\end{align}
where we take the coordinates of the second end-point $x_2^{\pm}$ as well as $x_1^+$ to be fixed. Recall that the contribution of the function $\tilde{\mathcal{F}}_n$ to the entropy is sub-leading although it contributes to the primary QNEC. Using the expression for $f'$ \eqref{Eq:sprimezero} and the entanglement entropy as a function of $x^-$, we can explicitly obtain the expression for $f(x^-)$.

\begin{figure}
    \subcaptionbox{}[0.475\textwidth]{\includegraphics[width=0.42\textwidth]{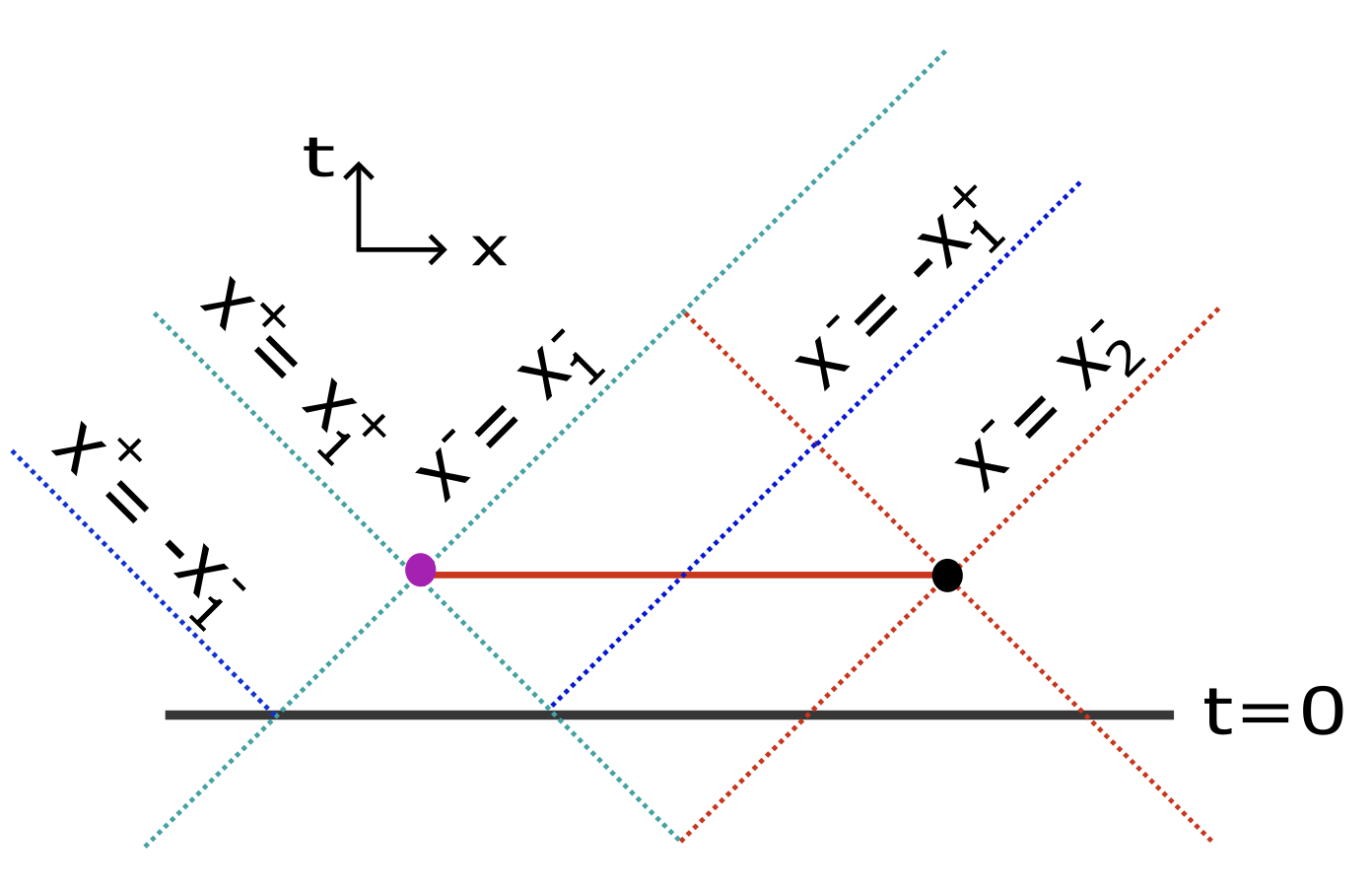}}
    \subcaptionbox{}[0.475\textwidth]{\includegraphics[width=0.45\textwidth]{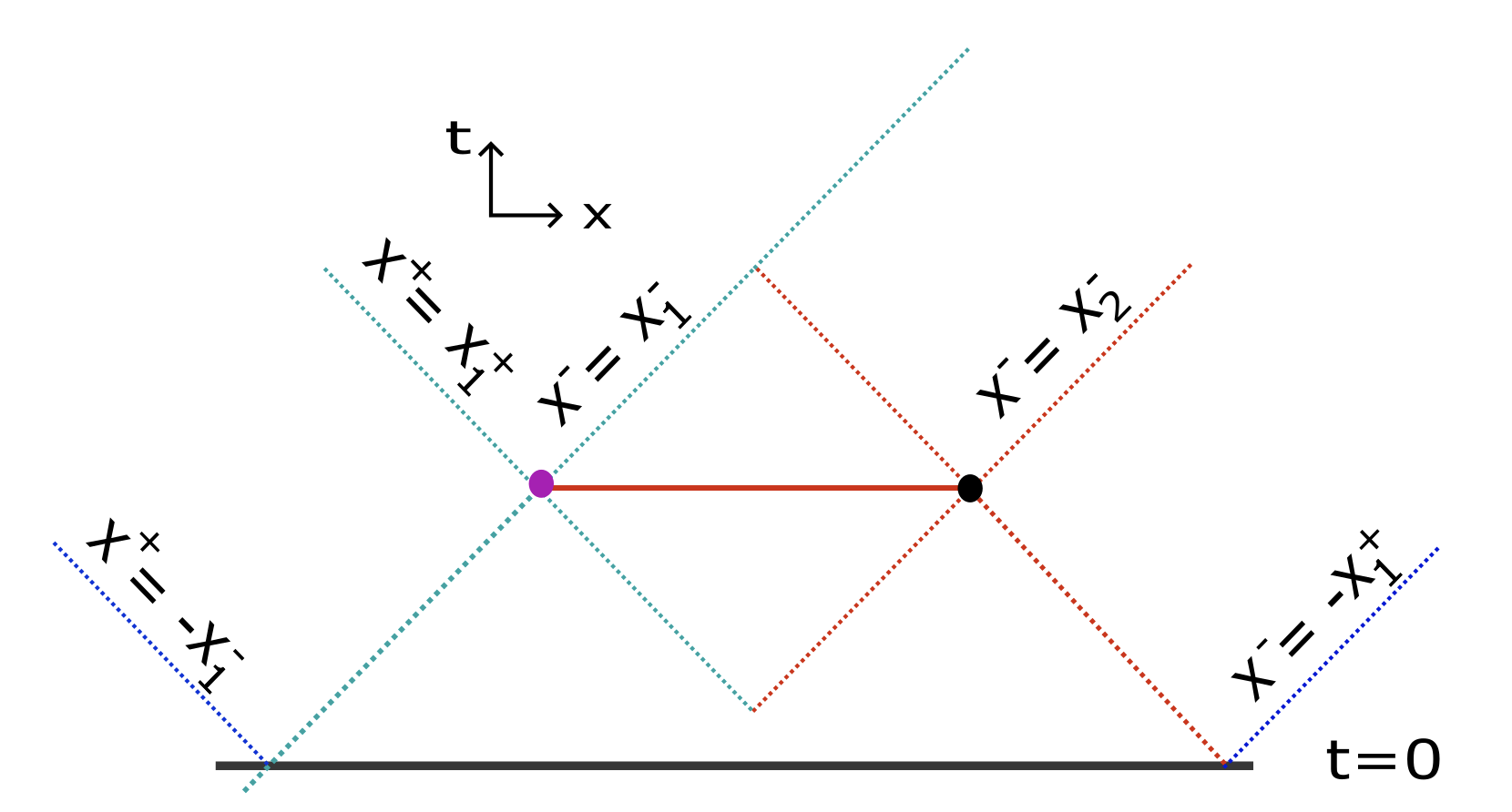}}
    \caption{Figure to illustrate the limits of the ANE integrals. The interval of interest is shown in red with end-points marked in magenta and black. The QNEC is evaluated at the magenta point. The ANE operators are integrated along the cyan null directions. The null coordinates should be integrated for $t>0$. This leads to a lower bound of $x^+=-x_1^-$ and $x^-=-x_1^+$. (a) The case when $x_2^-<-x_1^+$. For the global quench this is when $t<\frac l2$. (b) The case when $x_2^->-x_1^+$, corresponding to $t>\frac l2$ in the global quench.}
    \label{Fig:lightcone}
\end{figure}

From Fig.~\ref{Fig:lightcone} it is easy to see that the $x^-$ coordinate, for the null cut at the first end-point of the interval, ranges from 
\begin{equation}
    x^- \in \lB\frac l2-t,\infty\rb,
\end{equation} 
where the lower bound corresponds to the boundary at $t=0$. 

 Similarly, we can define $S(x^+)$, which is needed to define the map $\bar{f}(x^+)$ to a frame where the derivative of the entropy with respect to $x^+$ vanishes, by fixing the coordinates $x_2^\pm$ and $x_1^-$. In this case, we can again see from Fig. \ref{Fig:lightcone} that
\begin{equation}
    x^+ \in \lB-\frac l2-t,\infty\rb.
\end{equation}

 We find that
 \begin{equation}
     \bar{f}(x_2^+)\to\infty, \quad f(x_2^-)\to\infty
 \end{equation}
Thus the ANE operator in the $f$-frame,
 \begin{equation}
     \int\braket{T(f)}df,
 \end{equation}
 generates null deformations of the first end-point but doesn't affect the second end-point. By transforming coordinates, we find
\begin{align}
    \braket{T(f)}=&\ \left(\braket{T(x^-)}+\frac{c}{24 \pi}{\rm Sch}(f(x^-),x^-)\right)f'(x^-)^{-2} \nonumber \\
    =&\ \frac{1}{2\pi} \left(2\pi \braket{T(x^-)}-\partial_-^2S(x^-)+\frac{6}{c}(\partial_- S(x^-))^2\right)f'(x^-)^{-2},\\
    =&\ \frac{1}{2\pi} \mathcal{Q}_-(x^-) f'(x^-)^{-1}
\end{align}
where we have used \eqref{Eq:sprimezero} to obtain the second equality. 

To check whether our calculations and results satisfy the assumptions of the CF proof of QNEC, we need to check that the following is finite
\begin{align}
\braket{\mathcal{\tilde{P}_-}}\coloneqq\frac{1}{2\pi}\int_{\frac{l}{2}-t}^{\infty}dx^-\mathcal{Q}_-(x^-)f'(x^-)^{-1}.
    \label{Eq:anefinxm}
\end{align}
A similar analysis applies for the anti-holomorphic part, for which we need to check finiteness of
\begin{align}
\braket{\mathcal{\tilde{P}_+}}\coloneqq\frac{1}{2\pi}\int_{-\frac{l}{2}-t}^{\infty}dx^+\mathcal{Q}_+(x^+)\bar{f}'(x^+)^{-1}.
    \label{Eq:anefinxp}
\end{align}
Note that the derivatives that appear in the above integrals are non-negative, see \eqref{Eq:sprimezero}. From this it is clear that QNEC violation for all $x^\pm$ would imply that the averaged null energy is negative, thus violating the averaged null energy condition (ANEC) in the $f$-frame.

We can explicitly compute the primary QNEC $\mathcal{Q}_-$ for arbitrary $x^-_1$ and fixed $x_1^+,x_2^\pm$ with $t<\frac l2$. In this case the cross ratio $\eta$ remains approximately $0$ for the entire range of the integral and a theory independent analysis is therefore possible. Series expanding this expression in small $\epsilon$ we find that the integrand of $\braket{\mathcal{\tilde{P}}_-}$ is proportional to
\begin{equation}
    \frac{c}{6}+\partial_n\tilde{\mathcal{F}}'_n(0)\vert_{n=1},
\end{equation}
with the proportionality constant being a non-negative function of $x^-$. The integral can be readily computed using the small $\epsilon$ expansion as
\begin{equation}
    \braket{\mathcal{\tilde{P}_-}}=e^{\frac{-(l-4t)}{2\epsilon}}\frac{\epsilon^3}{\a\pi^3}\left(\frac{c}{6}+\partial_n\tilde{\mathcal{F}}'_n(0)\vert_{n=1}\right)
    \label{Eq:aneglobal}
\end{equation}
The finiteness of this integral implies that the CF proof should apply and the primary QNEC $\mathcal{Q}_-\geq0$ should be valid in this setup.
A violation of the QNEC bound \eqref{Eq:qnecboundglobal} implies that the ANEC is violated in the $f$-frame {where the} first null variation of the entanglement entropy {vanishes}. 

When $t>\frac l2$, the range of the integral \eqref{Eq:anefinxm} crosses $x^-=0$.  Similarly, the integral in the ANE \eqref{Eq:anefinxp} for the $+$ null direction crosses $x^+=0$ for both $t<\frac l2$ and $t> \frac l2$. Our analysis doesn't shed light on the validity of the CF proof for $\mathcal{Q}_{\pm}$ in these cases. This is because the CFT description for the quench breaks down when $x^\pm$ are comparable to the cutoff $\epsilon$. The cutoff is related to the correlation length of the critical lattice system and the CFT description is valid at scales much larger than the correlation length \cite{Calabrese:2005in}. Furthermore, the cross ratio $\eta$ as a function of $x^+$ varies from $0$ to $1$ along the range of the integral. As a function of $x^-$, when $t> \frac l2$, $\eta$ ranges from $2$ to $1$ along the range of the integral. The integral is then theory dependent, since we would need to know the function $\tilde{\mathcal{F}}_n$ explicitly for $\eta$ away from $0$ and $1$. It is thus not clear from our theory-independent CFT analysis if the CF proof in the $f$-frame and the primary QNEC apply to this setup in these cases.

We conclude this section by emphasising that the primary QNEC bounds \eqref{Eq:qnecboundglobal} should be thought of as imposing constraints on the set of allowed boundary states such that these states satisfy the ANEC in a conformally transformed frame, where the first variation of the entanglement entropy, with respect to the $x^-$ direction, vanishes. Since a violation of QNEC implies a violation of causality under modular evolution \cite{Balakrishnan:2017bjg}, we can also think of the bound \eqref{Eq:qnecboundglobal} on the four point function as a modular causality restriction on the set of boundary states. Similar bounds on the three point correlators of the theory were found by demanding that the ANEC be satisfied \cite{Hofman:2008ar}.

\section{Local joining quench}     \label{sec:local}
In this section we consider the QNEC in a local joining quench \cite{Calabrese:2007mtj}. Consider cutting a {$1+1$ dimensional} CFT at $x=0$ into two half lines $L$ and $R$. Prepare the half-line vacua on $L$ and $R$. The two halves are joined at time $t=0$. The state just before joining is simply the tensor product of the two half line vacua $\ket{0}_L \otimes \ket{0}_R$. The state after joining is evolved in time using the CFT Hamiltonian on the full line. Entanglement entropy can be calculated using a correlation function on the right half plane, similarly as described in the previous section.

\begin{figure}
    \centering
    \includegraphics[scale=0.65]{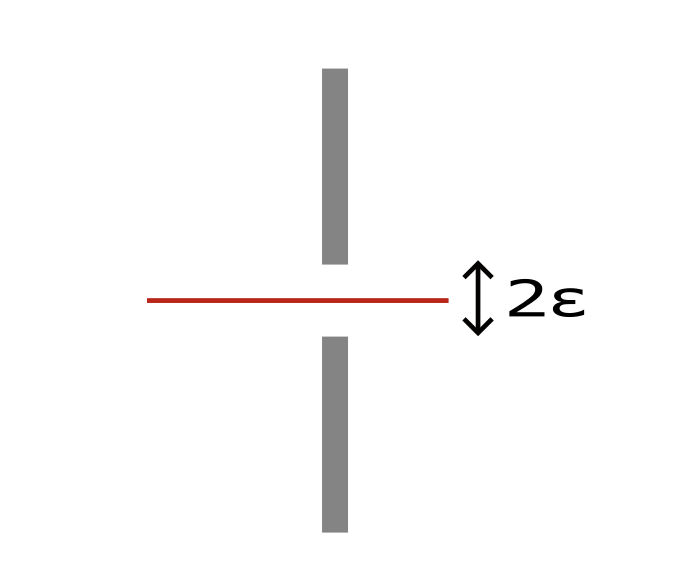}
    \caption{The geometry used to compute the Euclidean path integral. The vertical axis is imaginary time and the boundary conditions are the vacuum state. The two grey slits, running from Euclidean time $\tau=-\infty$ to $-\epsilon-i t$ and from $\epsilon-it$ to $\infty$, indicate that the left and right regions are decoupled. The red line indicates a slit along the subregion whose density matrix is being calculated. This figure is adapted from \cite{Calabrese:2007mtj}.}
    \label{fig:local-setup}
\end{figure}
 
 The reduced density matrix of an interval can be computed using an Euclidean path integral on the plane with two slits on the imaginary time axis at $x=0$ as shown in Fig.\ref{fig:local-setup}. One slit runs from Euclidean time $\tau=-\infty$ to $-\epsilon-i t$ and the other runs from $\epsilon-it$ to $\infty$. The slits indicate that the two half lines are decoupled. As in the previous section, we also have a branch cut along the interval of interest. This plane with slits can be mapped to the half plane ${\rm Re}(z)>0$ via
\begin{equation}
    z(w) = \frac{w}{\epsilon}+\sqrt{\frac{w^2}{\epsilon^2}+1}.
\end{equation}
The R\`{e}nyi entropies and hence the entanglement entropy can be obtained using the correlator \eqref{Eq:renyicorr} on the half plane. The cross-ratio is now
\begin{equation}
    \eta = \frac{(z_1 + \bar z_1)  (z_2 + \bar z_2)}{(\bar z_1 + z_2) (z_1 + \bar z_2)},
\end{equation}
and the entanglement entropy is
\begin{equation}
    S_A = \frac c6 \log\lB (z_1 - z_2) (\bar z_1 - \bar z_2)\eta \rB - \partial_n\tilde{\mathcal{F}}_n(\eta)\vert_{n=1} \ .
\end{equation}
This can again be analytically continued to obtain expressions in terms of the lightcone coordinates. {However, as noted for the global quench, we should first compute null derivatives and only then analytically continue to real time. Unlike the global quench, this subtlety in analytic continuation is important for the local quench {because of the branch cut in the square root map between the $z$ and $w$ coordinates.}} 
The energy-momentum tensor is computed using the Schwarzian derivative of the above map and is
\begin{equation}
    \braket{T_{\pm\pm}(x^\pm)} = \frac{c}{16 \pi}\frac{\epsilon^2}{({x^\pm}^2+\epsilon^2)^2}.
\end{equation}
Note that $\epsilon$ regulates the width of the energy-momentum tensor. In the $\epsilon \to 0$ limit this results in a $\delta$-function shockwave localized on the lightcone.

\subsection{QNEC}

\begin{figure}
    \centering
    \includegraphics[scale=0.35]{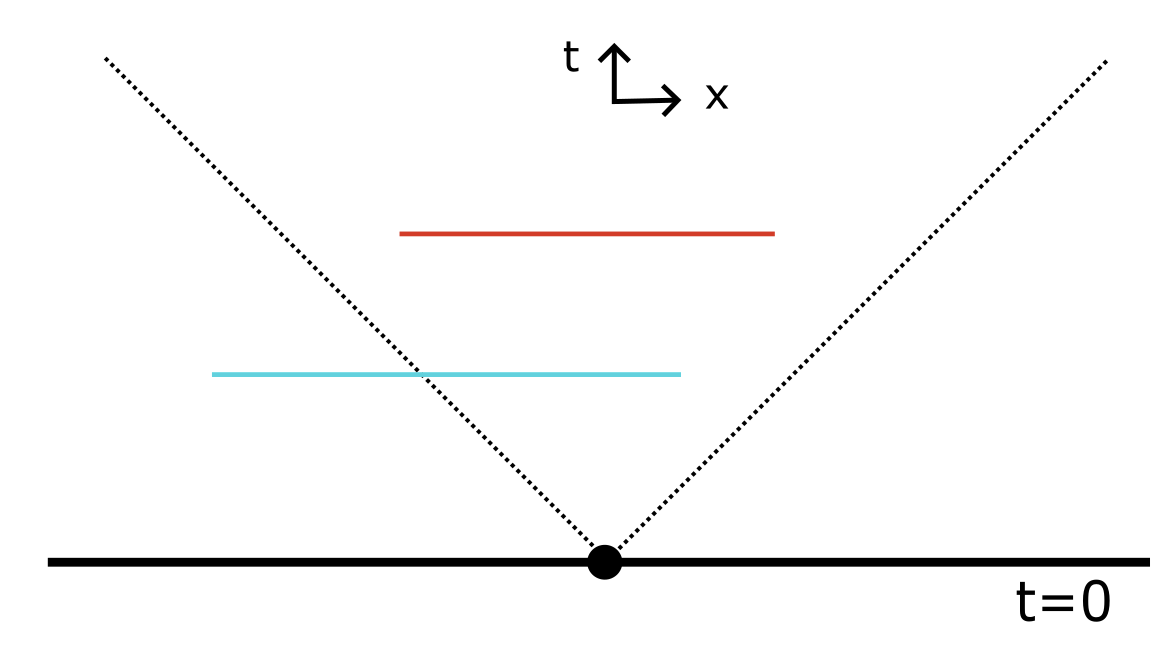}
    \caption{A plot to illustrate the entangling regions where a theory-independent primary QNEC can be calculated. The black circle indicates the point $x=0$ where the two half lines are joined. The dashed lines are the (regulated) shockwaves propagating along the null directions. Note that the red interval trivially satisfies the weak QNEC, whereas the blue interval leads to non-trivial bounds from the weak QNEC.}
    \label{Fig:localintervals}
\end{figure}

One can evaluate the QNEC using the above expressions for entropy and energy-momentum tensor. There are various different choices for the entangling interval. For example, the interval can include or exclude the point $x=0$ where the two halves are joined. We take the end-points of the interval to be
\begin{equation}			\label{eq_end_pts_local_quench}
	x_1^+ = t +l_1, \quad x_1^-  = t-l_1, \qquad x_2^+ = t + l_2, \quad x_2^- = t - l_2.
\end{equation}
Without loss of generality, we can assume that $l_2>l_1$. The cross-ratio $\eta$ varies depending on the choices for signs of the null coordinates. When $\eta$ is not close to zero or one, a theory-independent analysis is not possible. The two cases where a theory-independent analysis is possible are illustrated in Fig.~\ref{Fig:localintervals}.

If $x_{1,2}^\pm$ are all positive, i.e., when $t\pm l_2>0,t\pm l_1>0$, as indicated  by the red interval in Fig.~\ref{Fig:localintervals}, we find that the conformal cross ratio $\eta \approx 1+ \mathcal{O}(\epsilon^4)$. In this case, we can evaluate the derivatives of entanglement entropy in a theory-independent manner. We get,
\begin{align}
    \cQ_{1}^- =&\ \frac{\e^2}{4(t-l_1)^4}  \frac{2(t-l_1)(t-l_2)}{(t+l_1)(t+l_2)}\lb \frac c6 - \del_n\tilde\cF_n'(1)\vert_{n=1} \rb  ,    \\
    \cQ_{2}^- =&\ \frac{\e^2}{4(t-l_2)^4}  \frac{2(t-l_1)(t-l_2)}{(t+l_1)(t+l_2)}\lb \frac c6 - \del_n\tilde\cF_n'(1)\vert_{n=1} \rb  ,\\
    \cQ_{1,2}^+ =&\ \frac{\e^2(t+l_2)}{2(t-l_1)(t+l_1)^3(t-l_2)}\lb \frac c6 - \del_n\tilde\cF_n'(1)\vert_{n=1} \rb 
\end{align}
Imposing $\cQ_{1,2}^{\pm}\geq0$ leads to the same bound on $ \del_n\tilde\cF_n'(1)\vert_{n=1}$ as the global quench \eqref{Eq:qnecboundglobal}.
Note that the weak QNEC is always satisfied for this interval and is independent of $\tilde\cF_n$ at leading order.

The other case where a theory-independent analysis is possible is when $x_1^+<0$ and the other lightcone coordinates are all positive (blue interval in Fig. \ref{Fig:localintervals}). Note that since we have $l_2>l_1$, this case is only possible when $l_1<0$. The QNEC expressions for this case also involve $\lb \del_n\tilde\cF_n'(1)\vert_{n=1}\rb^2$ at the leading order
\begin{align}
    \cQ_1^- =&\ \Bigg(\frac c6 \frac{4 (t-l_1)^2 (t-l_2) (t+l_1)}{l_1(l_2-l_1)\e^2} - \frac{4(t-l_1)(t-l_2)(t+l_1)(l_1^2 - 3 l_1 l_2 + 3 l_1 t - l_2 t)}{l_1(l_2-l_1)(l_2+l_1)\e^2}\del_n\tilde\cF_n'(1)\vert_{n=1}     \nn  \\
    &\ - \frac{4(t+l_1)^2(t-l_2)^2}{(l_2+l_1)^2\e^2}\lb \del_n\tilde\cF_n'(1)\vert_{n=1}\rb^2 \Bigg)\frac{\e^2}{4(t-l_1)^4}.   
    \label{Eq:qnecmppp1}
\end{align}
It is easy to see that the above quadratic equation in $ \del_n\tilde\cF_n'(1)\vert_{n=1}$ is negative at $\del_n\tilde\cF_n'(1)\vert_{n=1} \to \pm \infty$. Then, $\cQ_1^-\geq 0$ requires that $\del_n\tilde\cF_n'(1)\vert_{n=1}$ is bounded by the roots of $\cQ^-_1=0$. We find that the larger root approaches its minimum value $\frac c6$ as $t\to -l_1$ and the smaller root approaches its maximum $-\frac c6$ as $t\to l_2$, leading to the following bounds
\begin{equation}
    \frac c3\geq\frac c6-\del_n\tilde\cF_n'(1)\vert_{n=1}\geq0.
    \label{Eq:qneclocalbound}
\end{equation}
The lower bound above is the same as that obtained in the global quench \eqref{Eq:qnecboundglobal}, whereas the upper bound is a new result for the local quench.
Similarly, we obtain
\begin{align}
   \cQ_1^+ =&\  \frac{\e^2}{4(t+l_1)^4}\Bigg(\frac{4 \del_n\tilde\cF_n'(1)\vert_{n=1} (l_1-l_2) (l_1+t)^3 (l_2-t) \left(l_1^2-3 l_1 l_2+3 l_1 t-l_2 t\right)}{l_1 \epsilon ^2 (l_1+l_2)^3 (l_1-t)} \\
  &\ - \frac{4 \lb\del_n\tilde\cF_n'(1)\vert_{n=1}\rb^2 (l_1-l_2)^2 (l_1+t)^4 (l_2-t)^2}{\epsilon ^2 (l_1+l_2)^4 (l_1-t)^2}+\frac{2 c (l_1-l_2) (l_1+t)^3 (l_2-t)}{3 l_1 \epsilon ^2 (l_1+l_2)^2}\Bigg),  \nn       \\
  \cQ_2^- =&\ \frac{\epsilon^2}{4(t-l_2)^4}\Bigg(\frac{8 c l_1 (l_1+t) (l_2-t)^3}{3 \epsilon ^2 (l_1-l_2) (l_1+l_2)^2} -\frac{16 l_1^2 \lb\del_n\tilde\cF_n'(1)\vert_{n=1}\rb^2 (l_1+t)^2 (l_2-t)^4}{\epsilon ^2 (l_1+l_2)^4 (l_1-t)^2} \nn \\
  &\ +\frac{16 l_1 \del_n\tilde\cF_n'(1)\vert_{n=1} (l_1+t) (l_2-t)^3 \left(l_1^2-3 l_1 l_2+3 l_1 t-l_2 t\right)}{\epsilon ^2 (l_1-l_2) (l_1+l_2)^3 (l_1-t)}\Bigg),     \\
  \cQ_2^+ =&\ \frac 4 {\e^2} \Bigg(\frac{2 c (l_1-l_2) (l_1+t)^3 (l_2-t)}{3 l_1 \epsilon ^2 (l_1+l_2)^2}-\frac{4 \lb\del_n\tilde\cF_n'(1)\vert_{n=1}\rb^2 (l_1-l_2)^2 (l_1+t)^4 (l_2-t)^2}{\epsilon ^2 (l_1+l_2)^4 (l_1-t)^2}\nn \\
  &\ +\frac{4 \del_n\tilde\cF_n'(1)\vert_{n=1} (l_1-l_2) (l_1+t)^3 (l_2-t) \left(l_1^2-3 l_1 l_2+3 l_1 t-l_2 t\right)}{l_1 \epsilon ^2 (l_1+l_2)^3 (l_1-t)}\Bigg).
\end{align}
We analyse the QNECs above in the same way as $\cQ_1^-$ \eqref{Eq:qnecmppp1} and obtain the same bounds \eqref{Eq:qneclocalbound}.

We also calculate the weak QNEC in this case. Unlike the global quench, we find that even the weak QNEC imposes bounds on $\del_n\tilde\cF_n'(1)\vert_{n=1}$. When $x_{1,2}^\pm$ are all positive, all the weak QNEC inequalities evaluate to the same form and are trivially satisfied,
\begin{align}
    Q_{1,2}^\pm =&\ \frac{c}{6(l_1-l_2)^2} >0.
\end{align}
For the other interval where a theory-independent analysis is possible, with $x_1^+<0,x_1^->0,x_2^\pm>0,$ we get the following,
\begin{align}
    Q_1^- =&\ \frac{c}{24} \left(\frac{1}{l_1^2}+\frac{4}{(l_1-l_2)^2}-\frac{4}{(l_1-t)^2}\right) + \frac{2 (l_1+t) (l_2-t)}{(l_1+l_2) (l_1-t)^3} \del_n\tilde\cF_n'(1)\vert_{n=1}  \nn    \\
    Q_1^+ =&\ \frac{c}{24} \left(\frac{1}{l_1^2}-\frac{4}{(l_1+l_2)^2}+\frac{4}{(l_1+t)^2}\right) + \frac{2 (l_1-l_2) (l_2-t)}{(l_1+l_2)^3 (l_1-t)} \del_n\tilde\cF_n'(1)\vert_{n=1}    \\
    Q_2^- =&\ \frac{1}{6 (l_1+l_2)^3}\lb \frac{c (l_1+l_2) \left(l_1^4-2 l_1^2 l_2^2+4 l_1 l_2 (l_2-t)^2+l_2^4\right)}{(l_1-l_2)^2 (l_2-t)^2}-\frac{24 l_1  (l_1+t)}{l_1-t} \del_n\tilde\cF_n'(1)\vert_{n=1} \rb    \nn   \\
    Q_2^+ =&\ \frac{c}{6 (l_2+t)^2}     \nn
\end{align}
The solution of, e.g., $Q_1^-=0$ for $\del_n\tilde\cF_n'(1)\vert_{n=1}$ is dimensionless. If we choose the dimensionless variables to be say $l_2/l_1$ and $t/l_1$, we further notice that the range of these coordinates is finite due to the choice of our entangling region.  By searching numerically over this compact range, we find that the strongest bounds from these inequalities in fact approach the same bounds (\ref{Eq:qneclocalbound}) as those from the primary QNEC. This happens for intervals where $t,l_2\to -l_1$, i.e., intervals where both end-points are close to the two shock-waves, but $x_1^+<0$ and $x_2^->0$. As expected, the weak QNEC bounds are always weaker than the bounds from the primary QNEC. 

The weak QNEC bounds are only obtained for the blue region in Fig.~\ref{Fig:localintervals}. This region crosses the shock-wave, and the red region that trivially satisfies the weak QNEC does not cross any shocks.

\subsection{Averaged null energy}

We expect that the post local quench state has finite relative entropy with respect to the CFT vacuum since it is in the identity sector -- the fusion of two identities can only produce identity and descendants \cite{Stephan:2011kcw}. It is easy to see that the ANE for the weak QNEC is finite whenever the integral doesn't cross $x^\pm =0$. Thus the CF proof should hold for $Q_-$ in the case with $x_1^-<0,x_1^+>0,x_2^\pm>0.$

{To understand if the primary QNEC holds}, we must perform a similar analysis as that performed in Sec.~\ref{ssec:checkaneglobal} and verify if the ANE is finite. {We need to} evaluate the following integrals (see \eqref{Eq:anefinxm}, \eqref{Eq:anefinxp})
\begin{align}
    \braket{\mathcal{\tilde{P}_-}}&\coloneqq\frac{1}{2\pi}\int_{-l_1-t}^{\infty}dx^-\mathcal{Q}_-(x^-)f'(x^-)^{-1}\\
    \braket{\mathcal{\tilde{P}_+}}&\coloneqq\frac{1}{2\pi}\int_{l_1-t}^{\infty}dx^+\mathcal{Q}_+(x^+)\bar{f}'(x^+)^{-1}.
\end{align}
In the cases where a theory independent analysis of QNEC is possible, we find that a theory-independent analysis of the ANE is not possible. Either the cross-ratio is not near 0 or 1, or the range of the integral crosses $x^\pm=0$ where the CFT description breaks down. Therefore, in the local quench we are unable to evaluate the ANEs relevant to the primary QNEC and check for their finiteness without invoking an explicit theory. This implies that we cannot establish that the primary QNEC must hold for this quench. It would be interesting to study QNEC and the ANE in this quench for some explicit theories.

\section{Discussion}    \label{sec:discussion}

\subsection*{CFT}

The quantum null energy condition has been proven in a variety of contexts \cite{Bousso:2015wca,Malik:2019dpg,Koeller:2015qmn,Balakrishnan:2017bjg}. QNEC was proven in general Poincar\'{e} invariant quantum field theories for null cuts, in states with finite averaged null energy (ANE) and finite relative entropy with respect to the vacuum state by Ceyhan and Faulkner \cite{Ceyhan:2018zfg}. Out of these two restrictions, finite relative entropy is essential for the QNEC to be well defined. 

In order to understand our results in the context of the Ceyhan-Faulkner proof, we have tried to examine if the setups studied here satisfy the two assumptions made in the proof. We have argued for the finiteness of relative entropies of the out-of-equilibrium states we study here, and have computed the ANE whenever it is well defined in a theory-independent way.

For the global quench we find that the primary QNEC imposes bounds on the theory dependent function $\partial_n\tilde{\mathcal{F}}_n$, and hence on the two point functions of twist fields on the UHP. In the case where a theory-independent analysis is possible \eqref{Eq:aneglobal}, we find that whenever the primary QNEC bounds are satisfied then $0<\braket{\tilde{\mathcal{P}}}<\infty$, and a violation of these bounds  implies a violation of the ANEC in a conformally transformed frame.  The ANEC has been proven for general quantum field theories \cite{Klinkhammer:1991ki,Kelly:2014mra,Faulkner:2016mzt,Hartman:2016lgu,Kravchuk:2018htv}, although the presence of boundaries needs some care \cite{Fewster:2006uf}. Our bounds can be thought of as additional restrictions on the allowed boundary states, arising from the ANEC and also from causality in modular time \cite{Balakrishnan:2017bjg}. 

As far as we are aware the function $\tilde{\mathcal{F}}_n$ has not been calculated explicitly for non-integer $n$ in any theory. For example, the function $\mathcal{F}_n$ which appears in the four point correlator on the plane has been calculated for integer $n$ for the Luttinger liquid \cite{Calabrese:2009ez}, however the analytic continuation to non-integer $n$ is not straightforward. Our bounds from QNEC place restrictions on the analytic continuations of {related functions $\tilde{\mathcal{F}}_n$} that are allowed for non-integer $n \gtrsim 1$. It would be interesting to explicitly compute the analytically continued $\tilde{\mathcal{F}}$ in some theories in consistency with the bounds we have derived.

For the local quench, we again find bounds on $\partial_n\tilde{\mathcal{F}}_n$ from primary QNEC. Furthermore, in this setup, the weak QNEC also imposes bounds on $\partial_n\tilde{\mathcal{F}}_n$, that are weaker than the primary QNEC but approach those bounds in a certain limit. We cannot calculate the ANE for the primary QNEC in a theory independent manner for any interval in the local quench, and therefore cannot comment on the validity of the CF proof to this setup.

\subsection*{Holography}

{We have performed our calculations of QNEC entirely within CFT, without assuming an Einstein gravity dual.} Our results will also hold in the large central charge and sparse spectrum limit{, i.e., when an Einstein dual exists for the CFT}. The holographic proof of QNEC \cite{Koeller:2015qmn} is applicable to the quenches considered here. Therefore, we {should} also understand the QNEC violations we find in the context of the holographic proof. 

{The bulk dual of the global quench, in the UHP coordinates, is simply vacuum AdS$_3$ with an end-of-the-world brane corresponding to the boundary at $t=0$.} The bulk dual for the post local quench state discussed here is a locally AdS$_3$ Ba\~nados geometry \cite{Banados:1998gg} with an end of the world brane. In the local quench the regime where we find non-trivial bounds corresponds to bulk geodesics that intersect the end of the world brane. This can be deduced from the entanglement entropy, which is different from the entropy for the connected geodesic in a Ba\~nados geometry. The holographic QNEC follows from bulk entanglement wedge nesting. As discussed in a footnote in \cite{Wall:2012uf}, consistency theorems (such as nesting) for extremal surfaces that compute holographic entanglement entropy can fail in spacetimes with boundaries at a finite distance. For example, violations of strong sub-additivity of holographic entanglement entropy have been found in cases where the bulk duals are at a finite radial cutoff \cite{Sanches:2016sxy,Grado-White:2020wlb}. Thus, our results do not contradict the holographic proof of QNEC. We also note that \cite{Banerjee:2024wtl} do not find any violations of QNEC for $T\bar T$-deformation of 2d CFTs, implying that the presence of finite boundaries does not necessarily imply violations of QNEC. {It will be interesting to relate the recent discussion of entanglement wedge nesting in the presence of end of the world branes in \cite{Saraswat:2025gxp} with our results for QNEC.}

\subsection*{Future directions}
In this paper we have considered the simplest local joining quench where two half-line vacua are connected suddenly. More general setups where the two half-lines have different temperatures have been studied \cite{Bernard:2016nci}. Interesting steady-state formation has been observed in such setups, which have also been studied holographically in \cite{Erdmenger:2017gdk}. It would be interesting to examine QNEC in these setups.

A R\`{e}nyi generalization of the QNEC \cite{Lashkari:2018nsl}, which uses the sandwiched R\`{e}nyi divergences, has been proven for free field theories \cite{Moosa:2020jwt,Roy:2022yzm}. However, it is still unclear if the R\`{e}nyi QNEC is valid more generally. The quenches studied in this work are a natural place to explicitly compute and test the  R\`{e}nyi QNEC in non-trivial out-of-equilibrium states using both holography and purely CFT techniques. 

Furthermore, using the map between the Temperley-Lieb and Virasoro algebras \cite{Koo:1993wz} it might be possible to understand the implications of QNEC directly on spin chains. This can likely shed some further light on the QNEC violations we find in this paper.

\subsection*{Acknowledgements}
We would like to thank Thomas Faulkner for valuable discussions and for pointing us to the correct ANE operator that should be checked for finiteness. We would also like to thank Aron Wall for  helpful correspondence. We are also grateful to Adwait Gaikwad, Sam van Leuven, Ayan Mukhopadhyay, and Ronak Soni for helpful discussions. The research of PR is supported by an NRF Freestanding Postdoctoral Fellowship. PR would like to thank Chennai Mathematical Institute for their generous hospitality during the initial stages of this work.

\providecommand{\href}[2]{#2}\begingroup\raggedright\endgroup

\end{document}